\documentclass[11pt, a4paper]{article}
\usepackage{jheppub}
\usepackage{graphicx}
\usepackage{epsfig,amsmath,amsthm}
\usepackage{epstopdf}

\newcommand{\be}{\begin{equation}}
\newcommand{\ee}{\end{equation}}
\newcommand{\bea}{\begin{eqnarray}}
\newcommand{\eea}{\end{eqnarray}}
\def\bse{\begin{subequations}}
\def\ese{\end{subequations}}

\def\IZ{\relax\ifmmode\hbox{Z\kern-.4em Z}\else{Z\kern-.4em Z}\fi}

\def\del{{\partial}}

\def\hphi{{\hat \phi}}

\def\hS{{\hat S}}  \def\hQ{{\hat Q}} \def\hx{{\hat x}}
 \def\hQ{{\hat Q}}
\def\hrho{{\hat \rho}}

 \def\co{{\cal O}}

  \def\eps{\epsilon}

 \def\hrho{{\hat \rho}}

\def\presub{\vspace{.5cm} \noindent}

\def\bi{\begin{itemize}} \def\ei{\end{itemize}}

\def\({\left(} \def\){\right)}
\def\[{\left[} \def\]{\right]}

\def\w{\omega}

\def\d{\partial}

\def\Om{\Omega}

\def\Gm{\Gamma}

\def\gaugeA{\tilde{\bf A}}
\def\gaugeD{\tilde{\boldsymbol{\rho}}}
\def\fieldA{{\bf A}}
\def\fieldD{\boldsymbol{\rho}}

\title{Radiation reaction at the level of the action}

\author{Ofek Birnholtz, Shahar Hadar and Barak Kol\\
{\it Racah Institute of Physics, Hebrew University, Jerusalem 91904, Israel} \\
{\tt ofek.birnholtz@mail.huji.ac.il}, {\tt shaharhadar@phys.huji.ac.il}, {\tt barak.kol@mail.huji.ac.il}
}

\abstract{The aim of this paper is to highlight a recently proposed method for the treatment of classical radiative effects, in particular radiation reaction, via effective field theory methods.
We emphasize important features of the method, and in particular the doubling of fields.
We apply the method to two simple systems: a mass--rope system and an electromagnetic charge--field system.
For the mass--rope system in 1+1 dimensions we derive a double-field effective action for the mass which describes a damped harmonic oscillator.
For the EM charge--field system, i.e. the system of an accelerating electric charge in 3+1\,d,  we show a reduction to a 1+1\,d radial system of an electric dipole source coupled to an electric dipole field (analogous to the mass coupled to the rope).
For this system we derive a double-field effective action, and reproduce in an analogous way the leading part of the Abraham--Lorentz--Dirac force.}

\begin{document}
\maketitle

\section{Introduction}
 \label{section:Intro}

An action formulation for radiation reaction (RR) was presented in \cite{BirnholtzHadarKol2013a}.
While \cite{BirnholtzHadarKol2013a} focused on the post-Newtonian approximation to the two body problem in Einstein's gravity, it stressed the method's generality, and presented detailed calculations also for radiation reaction of scalar and electromagnetic (EM) fields.
Following demand, this paper highlights the method by focusing on the well-known electromagnetic Abraham--Lorentz--Dirac (ALD) force \cite{Abraham,Lorentz1,Lorentz2,Dirac,ALD-Jackson,ALD-Rohrlich}.

The method offers certain advantages over the more standard approach. First, the action formulation allows the application of efficient tools including the elimination of fields through Feynman diagrams, and allows ready formulation and systematic computation of analogous quantities in non-linear theories such as gravity. Secondly, a single effective action would be seen to encode several ALD-like forces and its form would stress a connection between the source and target not seen in the usual force expression.

We start in this section by surveying the background and reviewing the general method and its key ingredients before proceeding to concrete demonstrations in the following sections.
In section \ref{section:mass--rope} the method is illustrated in a simple set-up -- a mass attached to an infinite rope. Section \ref{section:ALD} describes the application of the method to the leading non-relativistic ALD force (also known as the Abraham-Lorentz force), i.e. the force acting on an accelerating electric charge due to its own EM field.
The derivation has a strong analogy with the mass--rope system, especially after a reduction to a radial system through the use of spherical waves; many technicalities of the reduction are deferred to Appendix \ref{sec:gauge}, and may be skipped in a first reading.

\presub  {\bf Background}.
Feynman diagrams and effective field theories were first introduced in the context of quantum field theories, but they turn out to be applicable already in the context of classical (non-quantum) field theories.
One such theory is the Effective Field Theory (EFT) approach to general relativity (GR) introduced in \cite{GoldbergerRothstein1}, see also contributions in \cite{CLEFT-caged,NRG}.

Similarly, the development of an action method for dissipative systems was first developed in the early 60's in the context of quantum field theory, and is known as the Closed Time Path formalism \cite{CTP}.
The essential idea is to formally introduce a doubling of the fields which can account for dissipation.
Here too relevant non-quantum problems exist, such as radiation reaction, and hence it was natural to seek a theory for the non-quantum limit.
This was discussed in the context of the EFT approach to GR in \cite{GalleyEFT} and was promoted to general classical non-conservative systems in \cite{GalleyNonConservative}.
We also recommend comparing the mass--rope system presented here with the case of coupling a mass to infinitely many oscillators, presented in \cite{GalleyNonConservative}, and with the action formulations of a damped harmonic oscillator in \cite{Kosyakov:2007qc}.
Essential elements of the theory were reformulated in \cite{BirnholtzHadarKol2013a}.
This formulation was extended from 4d to general dimensions in \cite{BirnholtzHadar2013b}.

\subsection*{Brief review of method}

Here we present the tools generally, before demonstrating their use in the following sections.

{\bf Field doubling}. The theory in \cite{BirnholtzHadarKol2013a} was formulated for classical (non-quantum) field theories with dissipative effects, such as radiation reaction.
The source for the departure from the standard theory was identified to be a \emph{non-symmetric (or directed) propagator}, such as the retarded propagator.
For such theories the fields are doubled together with the sources, so that the propagator always connects a field and its double and thereby assigns a direction to it.
Here the term field should be understood to refer also to dynamical variables in mechanics (viewed as a field theory in 0+1 d).

A generic theory may be described by a field $\phi=\phi(x)$ together with its source $\rho=\rho(x)$ and the equation of motion for $\phi$, $0=EOM_\phi(x)$, where the subscript notation readily generalizes to a multi-field set-up.
The double field action is given by \cite{BirnholtzHadarKol2013a}  \be
\hS \[ \phi,\hphi;\, \rho,\hrho \] :=
	\int d^d x\, \[ EOM_\phi(x)\,\hphi(x)
	+\int d^d y\, \frac{\delta EOM_\phi (y) }{\delta \rho(x)} \hrho(x)\, \phi(y) \],
\label{doubled-action}
\ee
where $\hphi$ is a doubled auxiliary field, $\hrho$ is the corresponding source, and $d^dx \equiv d^{d-1} {\bf x}\, dt$ is a space-time volume element.
The action is best described in terms of functional variations, but in our examples it will amount to simple concrete expressions shown in the following sections.
The expression is constructed such that $\delta \hS/\delta \hphi = EOM_\phi$, namely that the equation of motion with respect to $\hphi$ reproduces the original $EOM_\phi$.
The first part introduces terms of the form $\phi(x)\, \hphi(x)$, $\rho(x)\, \hphi(x)$, while the second part introduces terms of the form $\hrho(x)\, \phi(x)$.

This formulation is applicable to arbitrary equations of motion, not necessarily associated with an action.
When they \emph{can} be derived from an action $S=S[\phi,\rho]$,  such as in the case of radiation reaction, namely if $EOM_\phi=\frac{\delta S}{\delta \phi}$ for some $S$, the double field action is
\be
\hS \[ \phi,\hphi;\, \rho,\hrho \] :=
	\int d^d x\, \[ \frac{\delta S[\phi,\rho]}{\delta \phi(x)}\,\hphi(x)
	+\frac{\delta S[\phi,\rho]}{\delta \rho(x)}\, \hrho(x)\,  \] ~.
\label{doubled-action2}
\ee
Here $\hphi,\, \hrho$ can be assigned the following meaning: $\hphi$ is the linearized field perturbation, sourced by reverse (advanced) propagation from a generic source $\hrho$.
They correspond to the Keldysh basis of the Closed Time Path formulation and $\hS$ is a linearization of the CTP action with respect to $\hphi$.
We remark that this procedure determines the action explicitly, and in fact gives a recipe for the function $K$ mentioned in \cite{GalleyNonConservative}.
It also generalizes the specific field doubling method (using complex conjugation) of \cite{Kosyakov:2007qc}.

{\bf Zone separation}. The EFT approach is characterized by a hierarchy of scales leading to multiple zones (and a matched asymptotic expansion).
In particular when the velocities of the sources are small with respect to the velocity of outgoing waves, such as in the Post-Newtonian approximation to GR, one defines a system zone and a radiation zone.

{\bf Elimination}. Given $\hS$, one may proceed to eliminate fields through Feynman diagrams. \emph{Radiation sources $Q[x]$} are defined diagrammatically by
\bea
-Q[x] :=
\parbox{10mm} {\includegraphics[scale=0.3]{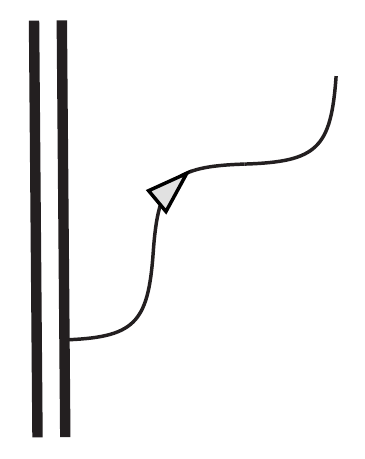}}
\label{def:Q}
\eea
where the double heavy line denotes the whole system zone.
Their definition involves the elimination of the system zone field and matching, see \cite{BirnholtzHadarKol2013a}  eq. (2.49), but this shall mostly not be required in the present paper.
The doubled radiation sources are given by
\be
 \hQ[x, \hx] = \int dt\, \frac{\delta Q[x]}{\delta x^i(t)}\, \hx^i(t) \, \, . \label{def:Qhat}
 \ee
which is the linearized perturbation of the source and as such does not require the computation of new diagrams beyond (\ref{def:Q}).
Here too this general formula will be seen to reduce to simple expressions in the examples.

Next the \emph{radiation reaction effective action} $\hS_{RR}$ may be computed through elimination of the radiation zone, see \cite{BirnholtzHadarKol2013a} eq. (2.53).
This action describes the effects of dissipation and radiation reaction.

{\bf Gauge invariant spherical waves in the radiation zone}. When the spatial dimension $D>1$ (as in the ALD problem) each zone respects an enhanced symmetry and corresponding field variables and gauge should be chosen:
\bi
\item In the system zone time-independence (stationarity) is an approximate symmetry, due to the slow velocity assumption, and hence non-Relativistic fields (and a compatible gauge) should be used \cite{NRG}.
\item In the radiation zone the system appears point-like and hence it was recognized in \cite{BirnholtzHadarKol2013a} that this zone is spherically symmetric and accordingly gauge invariant spherical waves (see \cite{AsninKol}) should be chosen as the field variables.
\ei

{\bf Matching within action}. Subsection 2.3 of \cite{BirnholtzHadarKol2013a} demonstrated how the matching equations which are a well-known, yet sometime sticky element of the effective field theory approach, can be promoted to the level of the action.
This is achieved by introducing an action coupling between the two zones, whereby the matching equations become ordinary equations of motion with respect to novel field variables.
These are termed two-way multipoles, and reside in the overlap region.
This is an important ingredient of our method, but is not used directly in this paper.

\section{Mass--rope system}
\label{section:mass--rope}

\begin{figure}[t!]
\centering \noindent
\includegraphics[scale=0.5]{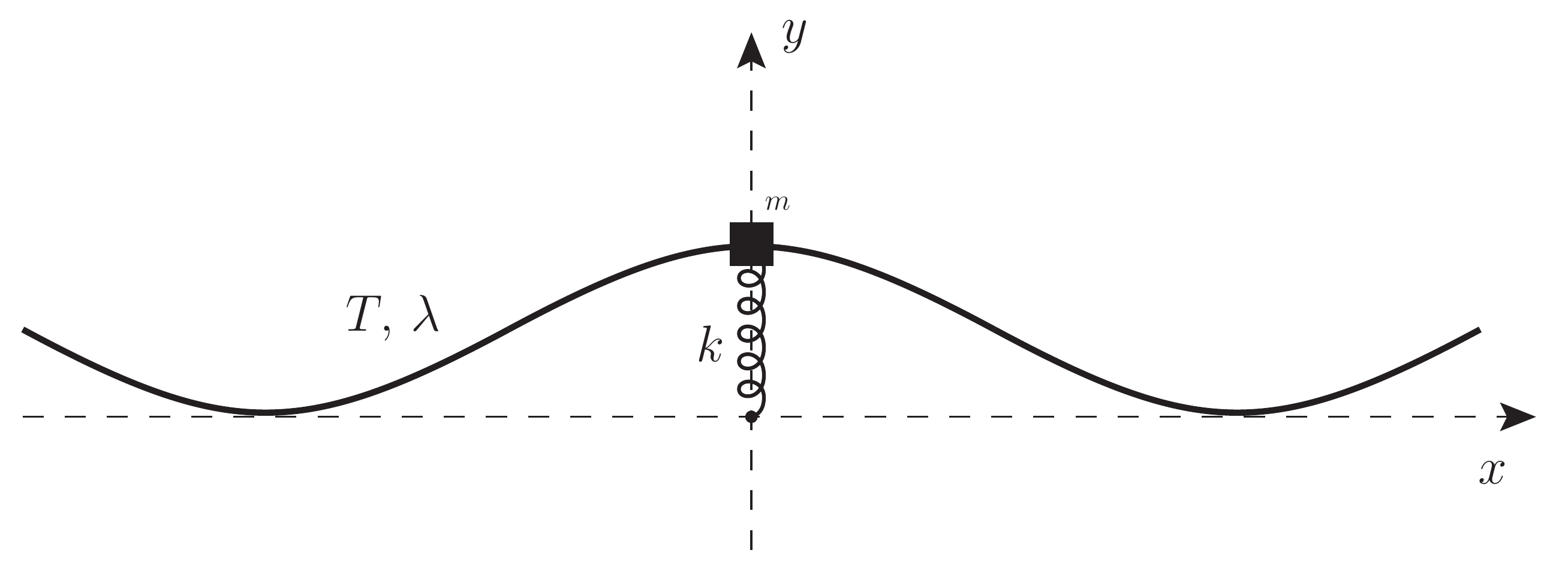}
\caption[]{The mass--rope system.}
 \label{fig:mass-rope}
\end{figure}

As a first example we consider a system shown in fig. \ref{fig:mass-rope} composed of a rope stretched along the $x$ axis with linear mass density $\lambda$ and tension $T$ (hence the velocity of waves along it is $c := \sqrt{T/\lambda}$).
Its displacement along the $y$ axis is given by $\phi(x)$.
The rope is attached at $x=0$ to a point mass $m$ which is free to move along the $y$ axis and is connected to the origin through a spring with spring constant $k$.
We assume there are no incoming waves from infinity.
Our goal is to find an effective action describing the evolution of the point mass in time \emph{without direct reference} to the field -- that is, after the field's elimination.
The eliminated field carries waves, and therefore energy, away from the mass; we show directly how this elimination gives a generalized (non-conservative) effective action describing the dynamics of the mass, which is that of a damped harmonic oscillator.

The action describing the system (mass+rope) is given by
\be
S \, = \,
	\frac{1}{2} \int\!\! dt \left[ m \dot{y}^2 - k y^2 \right]
	+ \frac{T}{2}\!\! \int\!\! dt dx \, \left( \d \phi \right)^2
	-  \int\!\! dt  \,Q \left[  \phi(0) - y  \right]	 \, \, ,
\label{mass rope action}
\ee
where $\left( \d \phi \right)^2 = \frac{1}{c^2}\dot{\phi}^2-(\d_x \phi)^2$ and $\dot{y} := \d_{t} y$.
The action is composed of the mass's action, a dynamical rope term, and a coupling term.
This coupling enforces the boundary condition $y \equiv \phi(x=0)$ at the level of the action and introduces the Lagrange multiplier $Q(t)$, where the choice of notation will be explained later.

We wish to obtain a long-distance effective action for the rope, and for that purpose we analyze the equations of motion.
Varying the action with respect to $\phi$ and $Q$ we find
\bea
\square \phi &=& -\frac{Q}{T}\, \delta(x) \label{EOM1} \, \, , \\
y &=& \phi(0) \label{EOM2} \, \, ,
\label{eom full theory}
\eea
where $\Box :=  \frac{1}{c^2} \d^2_t \, - \d^2_x$.
Analysis of (\ref{EOM1}) near $x=0$ gives $\left[ \d_x \phi \right]=Q/T$ where $\left[ \d_x \phi \right]:=(\d_x \phi)|_{0^+}-(\d_x \phi)|_{0^-}$ denotes the jump at the origin.
On the other hand the equations of motion and the outgoing wave condition imply
\be
\phi(x,t) =
	\left\{ \begin{array}{c}
		y(t - x/c)  \qquad x>0 \\
		y(t + x/c)  \qquad  x<0
	\end{array} \right.
\label{full theory radiation}
\ee
and hence $\left[ \d_x \phi \right] = -2 \dot{y}/c$.
Altogether we obtain
\be
\frac{Q}{T} = \left[ \d_x \phi \right] = -\frac{2}{c} \, \dot{y} \, \, .
\label{lambda1}
\ee

At this point we can substitute (\ref{lambda1}) in the eq. of motion for $y$ and obtain a damped harmonic oscillator.
Yet, here we wish to demonstrate how such a dissipative system may be described by a (double field) action.
For that purpose we proceed and find that the following action implies the same equations of motion\footnote
{ 	
In fact there is a slight difference: the effective action (\ref{mass rope far region effective action}) implies $\dot{\phi}(0)=\dot{y}$ which is not exactly the same as (\ref{EOM2}).
}	
including (\ref{lambda1})
\be
S \, = \,
	\frac{1}{2} \int\!\! dt \left[ m \dot{y}^2 - k y^2 \right]
	+ \frac{T}{2}\!\! \int\!\! dt dx \, \left( \d \phi \right)^2
	+ 2\, Z \!\! \int\!\! dt \, \dot{y} \,  \phi(0)		 \, \, ,
\label{mass rope far region effective action}
\ee
where  $Z:= \sqrt{T \lambda}$ is the rope's impedance.
We interpret $S$ as a long-distance effective action for $\phi$ (as it incorporates the solution (\ref{full theory radiation}) including the asymptotic boundary conditions).
Comparing (\ref{mass rope far region effective action}) with a standard origin source term $S_{\mathrm{int}} = -\int \! \phi(0,t) \, Q(t) dt \, \,$ justifies the notation $Q$.
In fact, $Q$ is the force exerted on the mass by the rope.

We double the field $\phi$ as well as its source $y$ in (\ref{mass rope far region effective action}) as described in (\ref{doubled-action2}) and find the double field action
\be
\hat{S} \, = \, T \int dt\, dx\, \d \phi\, \d \hat{\phi} + 2\, Z \!\! \int\!\! dt \[ \dot{y} \,  \hat{\phi}(0) + \dot{\hat{y}}\, \phi(0) \] \, , \,
\label{mass rope action frequency domain}
\ee
Transforming to frequency domain with the convention
$\phi(t) = \!\int\! \frac{d\w}{2\pi}\, \phi(\w)\, e^{-i \w t}$,
complex conjugate fields appear such as $\phi^*(\w)$ and we obtain the Feynman rules.
The directed propagator from $\hat{\phi}^*$ to $\phi$ is
\begin{align}
\label{Feynman Rule Propagator}
\parbox{20mm} {\includegraphics[scale=0.5]{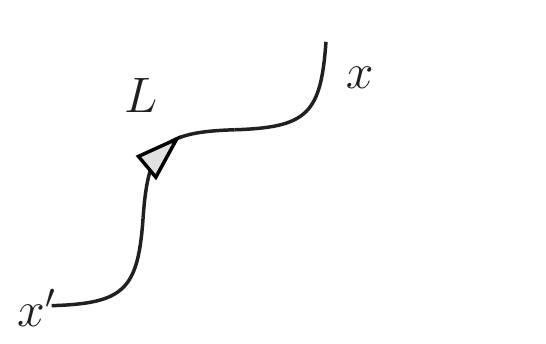}}
=G_{\w}(x',x) \, = \, - c\,  \frac{e^ {i \frac{\w}{c} \, |x-x'|}}{i\, \w \, Z}  ~,
\end{align}
and the sources for the fields $\hat{\phi}^*,\, \phi$ are (compare (\ref{def:Q}),(\ref{def:Qhat}))
\begin{align}
\label{Feynman Rule Vertex}
-Q_\w \equiv \parbox{20mm} {\includegraphics[scale=0.3]{DiagsGeneralRulesVertex.pdf}}
\!\!\!\!\!\!\!\!\!\!= - i \w \,  Z \, y_{\omega}~~~~~ ,~~\,
-\hat{Q}^*_\w \equiv \parbox{20mm} {\includegraphics[scale=0.3]{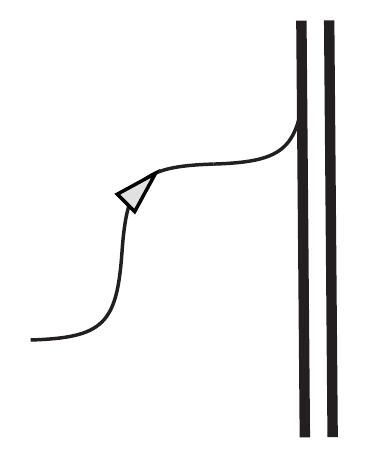}}
\!\!\!\!\!\!\!\!\!\!=i  \w \, Z \, \hat{y}^{*}_{\omega}~~ .~~~~~~
\end{align}
Similar Feynman rules hold for $\phi^*,\, \hat{\phi}$.

Now that we have the Feynman rules we can proceed to compute the outgoing radiation and the radiation reaction effective action.
Radiation away from the source, for $x>0$, is given by
\bea
\phi_\w(x) = \parbox{20mm}{\includegraphics[scale=0.5]{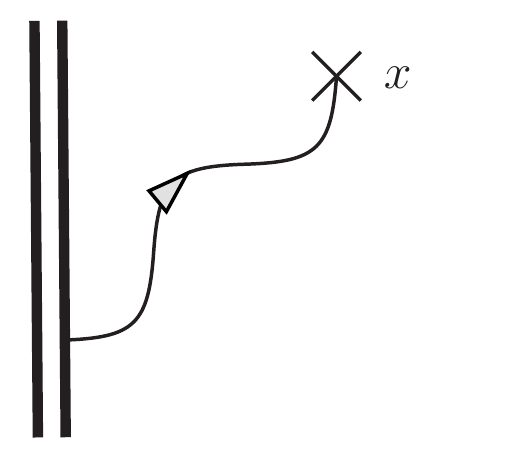}} = c \, y_{\w} \, e^{ i \w x /c} ~ ,
\label{rad}
\eea
which in the time domain becomes
\be
\phi(t,x) = \int \frac{d \w}{2 \pi}\,  y_\w \, e^{ - i \w (t -  \frac{x}{c})} = y(t -  \frac{x}{c})~ ,
\label{rad solution time}
\ee
reproducing (\ref{full theory radiation}).

The \emph{radiation reaction effective action} is found by eliminating the field $\phi$
\bea
\hat{S}_{RR} &=& \parbox{20mm}{\includegraphics[scale=0.5]{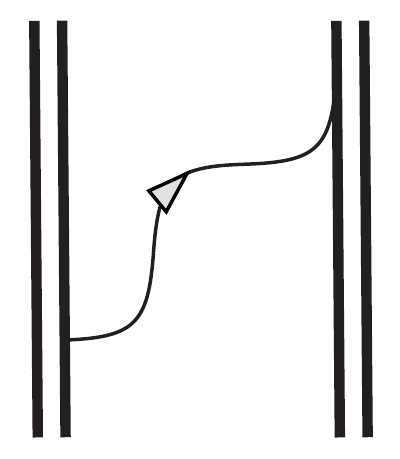}} ~ + ~ c.c. ~
	= \, \int \frac{d \w}{2 \pi} \, \, \hat{Q}^{*}_{\w} \, G_{\w}(0,0) \, Q_{\w} \, + c.c. \,
		= \nonumber \\
		&=& \int \frac{d \w}{2 \pi}  \, i \w \,  Z \, \hat{y}^{*}_{\w} \, y_{\w}
			~+ ~ c.c.
	=  \, -2\, Z \int \hat{y} \, \dot{y} \,  dt ~ .
\label{RR effective action}
\eea
Thus the full generalized effective action for the mass becomes
\be
\hat{S}_{tot} \, = \,	\hat{y} \frac{\delta}{\delta y }\left\{ \frac{1}{2} \int\!\! dt \left[ m \dot{y}^2 - k y^2\right]\right\}
	- 2 Z \int \hat{y} \d_t y \,  dt
\label{final action rope}
\ee
and by taking the Euler--Lagrange equation with respect to $\hat{y}$ we obtain
\be
m \ddot{y} = -k y - 2 Z \dot{y} ~ ,
\label{EOM for mass on rope}
\ee
which is the equation of a damped harmonic oscillator, as expected.

We remark that the $1+1$ mass--rope system can be regarded as a specific case of a scalar point charge coupled to a scalar field in $d=D+1$ dimensions, and thus falls under the general considerations of \cite{BirnholtzHadar2013b}, for $d=2$. Excluding relativistic, retardation and higher-multipole effects (irrelevant in our 1 spatial dimension), we recover the form given there for the radiation-reaction force (eq. 2.50), with $G q^2 = Z$, and a factor of 2 because the rope has two sides.

\section{Electromagnetic charge--field system}
\label{section:ALD}

The physical problem of the self-force (radiation-reaction force) on an accelerating electric charge has been treated for over 100 years \cite{Abraham,Lorentz1,Lorentz2,Dirac,ALD-Jackson,ALD-Rohrlich} in different methods.
We wish to show how in its simplest form and to leading order, the Abraham--Lorentz force can be easily found in essentially the same method as in the mass--rope system.
We start likewise with the standard Maxwell electromagnetic action\footnote
{	
We use Gaussian units, the speed of light $c=1$, and the metric signature is  $(+,-,-,-)$.
}	
\be
\label{MaxwellEMAction}
S_{full}=\!-\frac{1}{16\pi}\!\! \int\!\! d^4 x \,F_{\mu\nu} F^{\mu\nu} - \!\!\int\!\! d^4 x\, A_{\mu} J^{\mu},
\ee
where the field of the rope $\phi$ has been replaced by the EM field $A^\mu$ (with $F_{\mu\nu} \!=\! \d_\nu A_\mu \!-\! \d_\mu A_\nu$)\footnote
{	
As usual $F_{\mu\nu}$ encodes the electric and magnetic fields through $E_i=F_{0i}$, $B_i=-\frac{1}{2}\eps_{ijk}F_{jk}$.
}	
and the mass on a spring is replaced by a point-charge $q$ with trajectory ${\bf x}_p(t)$ and current density $J^\mu=q\frac{dx^\mu_p}{d\tau}\delta({\bf x} - {\bf x}_p)$.

As a 3+1\,d problem, this appears more complicated than the mass-rope system.
However, in the radiation zone the system appears point-like and the problem becomes spherically symmetric.
As dissipation and reaction are related to the waves propagating to spatial infinity, we can reduce the problem using the spherical symmetry  to an effective 1+1 dimensional ($r,t$) system, which can be treated analogously to the mass--rope system.
Physically, the reduction amounts to working with spherical wave variables which are described in appendix \ref{sec:gauge}.
For the purpose of the leading non-relativistic ALD force it suffices to consider the electric dipole ($\ell=1$) sector.
We denote the corresponding field variable by $\fieldA=\fieldA(r,t)$ and the source by $\fieldD=\fieldD(r,t)$, both defined in Appendix \ref{sec:gauge}.
In this sector the action, reduced to 1+1 dimensions, is given by (\ref{EM Action leading2}). In the time domain it becomes
\be
S= \int \!\! dt \! \int\!\! dr
	\left[
		- \frac{r^4}{12} \fieldA \!\cdot\! \Box \fieldA
		- \fieldA \!\cdot\! \fieldD
	\right],
\label{EM Action leading}
\ee
where now $\Box :=\d_t^2 - \d_r^2 - \frac{4}{r}\d_r$.
This field content is very similar to that of the rope: the coordinate $r$ replaces $x$, the field $\fieldA(r)$ replaces the field $\phi(x)$. As in (\ref{mass rope action}) the action has both a kinetic field term and a source-coupling term, though their form here differs.
The double-field action is found using (\ref{doubled-action2}) as in the mass--rope system, and is given by
\be
\hS = \int \!\! dt \! \int\!\! dr
	\left[
		- \frac{r^4}{6} \hat{\fieldA} \!\cdot\! \Box \fieldA
		- \( \hat{\fieldA} \!\cdot\! \fieldD + \fieldA \!\cdot\! \hat{\fieldD} \)
	\right].
\label{EM hatted Action leading}
\ee

The method thus proceeds similarly to find the Feynman propagator for the field and the expressions for the source vertices - non-hatted and hatted.
To obtain the propagator for $ \fieldA$ we transform to the $\w$ frequency domain and consider the homogenous part of the field equation (\ref{EOM PhiS}), which in dimensionless variables $x:= \w r$, becomes
\be
[ \del_x^2 + \frac{4}{x}\del_x + 1]\, \tilde{b}_{3/2}=0.
\label{Modified Bessel equation}
\ee
Its solutions are the origin-normalized Bessel functions
\be
\tilde{b}_{3/2} := \Gm(\frac{5}{2}) \, 2^{3/2}\, \frac{B_{3/2}\alpha(x)}{x^{3/2}}~ ,
\ee
where $B \equiv \{J,Y,H^\pm\}$ includes the Bessel functions $J,Y$ and the Hankel functions, $H^\pm=J \pm i\, Y$.
The origin-normalized Bessel $\tilde{j}_{3/2} = 1 + \co\(x^2\)$ is smooth at the origin and $\tilde{h}^+_{3/2} = -\frac{3}{x^2}\, e^{i\,x} \( 1 + \co\(\frac1x\) \)$ is an outgoing wave.
More details on these Bessel functions are given in Appendix \ref{sec:Bessel}.

Thus \emph{the propagator for spherical waves is}
\begin{align}
\label{Feynman Rule Propagator}
\parbox{20mm} {\includegraphics[scale=0.5]{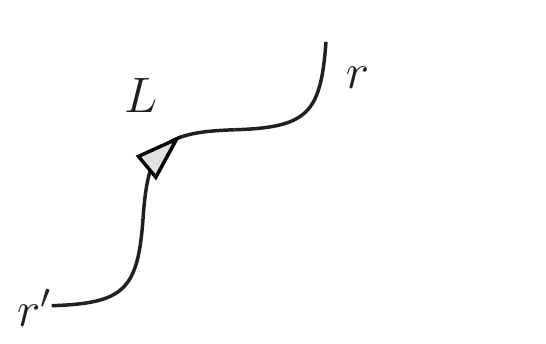}}
=G(r',r) \, = \, -\frac{2i\w^3}{3} \, \tilde{j}_{3/2}(\w r_1) \, \tilde{h}^+_{3/2}(\w r_2) \,;\\
r_1:=\text{min}\{r',r\},\,\,\,r_2:=\text{max}\{r',r\}.	\nonumber
\end{align}

In this sector, the source in the radiation zone is nothing but the electric dipole $ {\bf Q} \equiv {\bf D}$. Its form is identified through matching (\ref{def:Q}) the full theory with the radiation zone \be
\fieldA_{full}(r)
	\!=\! \int\!\! dr' \fieldD(r') \!\! \( \frac{-2i\w^{3}}{3}  \tilde{j}_{3/2}(\w r') \tilde{h}^+_{3/2}(\w r) \)
	\!\! =  {\bf Q} \frac{-2i\w^{3}}{3} \tilde{h}^+_{3/2}(\w r)
	\!=\! \fieldA_{rad}(r)~. ~~~
\label{EM scalar wavefunction at radiation zone1}
\ee
Using (\ref{def:Q},\ref{def:Qhat},\ref{Modified Bessel equation}, \ref{EM inverse sources}, \ref{EM source scalar Phi}) and integration by parts, we read from (\ref{EM scalar wavefunction at radiation zone1}) the electric dipole source vertex ${\bf Q}$ at leading non-relativistic order
\bea
\label{Feynman Rule Vertex}
\parbox{10mm} {\includegraphics[scale=0.3]{DiagsGeneralRulesVertex.pdf}}
&=& {\bf Q}\!=\!\!\!\int\!\!dr' \tilde{j}_{3/2}(\w r') \fieldD (r')
\!=\!-\frac{q_\w}{2}\!\!\int\!\! d^3 x'  \tilde{j}_{3/2}(\w r') {\bf n}' \left[r'^2 \delta({\bf x}' - {\bf x}_p) \right]'
	= q {\bf x}_p + \dots
	\equiv {\bf D}
\nonumber\\
\parbox{10mm} {\includegraphics[scale=0.3]{DiagsGeneralRulesVertexHat.pdf}}
&=&\hat{\bf Q} = \frac{\delta {\bf Q}}{\delta x^{i}}\hat{x}^{i}	= q \hat{\bf x}_p  + \dots
	\equiv \hat{\bf D}
\eea
We remark that this is merely the static electric dipole moment ${\bf D}$ of the source, where the ellipsis denote relativistic corrections.

With the Feynman rules at hand we can proceed to determine the outgoing radiation and the radiation reaction effective action.
Radiation away from the source is given diagrammatically by (compare \ref{rad})
\bea
\fieldA(r)=
\parbox{20mm}{\includegraphics[scale=0.5]{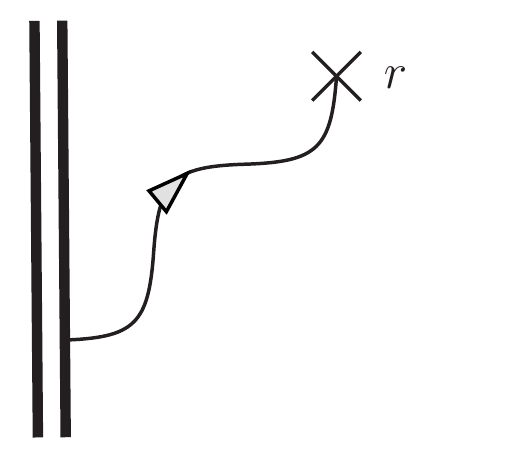}}
= 	-{\bf Q}\, G(0,r)
=-\frac{2i\,\w \, {\bf Q}\, e^{i \w\, r}}{r^2}~ , ~~~
\label{rad EM}
\label{Radiation EM using feynman scalar}
\eea
where we have used (\ref{Feynman Rule Propagator},\ref{Feynman Rule Vertex},\ref{Bessel H asymptotic2}). In the time domain, we find
\be
\fieldA ({\bf x},t)=\frac{2}{r}\, \d_t {\bf Q}(t-r) ~.
\label{radiation A}
\ee
The EM radiation reaction effective action is
\bea
\hS_{EM} =
\parbox{20mm}{\includegraphics[scale=0.5]{ActionDiagScalarL.pdf}} ~ + ~ c.c. ~
	&=& \, \frac{1}{2} \int \! \frac{d \w}{2 \pi} ~ \hat{{\bf Q}}^{*} \, G(0,0) \, {\bf Q} \, + c.c. \, \nonumber\\
&=&\frac{2}{3}\int\!\!dt \, \hat{\bf Q} \cdot \d_t^{3}{\bf Q}
\label{RR effective action EM general dipole}
\eea
where the propagator was evaluated at $r=r'=0$, as in (\ref{RR effective action}), and we regulated $ \tilde{h}^+(0)  \to \tilde{j}(0)=1$.
In the process of computing $\hS_{EM}$ the fields were eliminated and only the particle's dipole remains.
Note that the third time derivative originates from the $\w^3$ term in (\ref{Feynman Rule Propagator}), which in turn originates from the behavior of the Bessel function near its origin.

For a single charge we substitute (\ref{Feynman Rule Vertex}) in (\ref{RR effective action EM general dipole}) to obtain \be
\hS_{EM} = \frac{2}{3}q^2\!\!\int\!\!dt \, \hat{\bf x} \cdot \d_t^{3}{\bf x}   ~.
\label{RR effective action EM}
\ee
This can now be used to find the radiation-reaction (self) force, similarly to (\ref{final action rope},\ref{EOM for mass on rope}), through the Euler-Lagrange equation with respect to $\hat{\bf x}$
\be
{\bf F}_{RR} = \frac{2}{3} \, q^2 \dddot{\bf x}.
\label{EOM for charge ALD}
\ee
This of course matches the Abraham--Lorentz result \cite{Abraham,Lorentz1,Lorentz2}, and is the leading order term in the fully relativistic result of Dirac \cite{Dirac} (given in similar form in \cite{BirnholtzHadarKol2013a}, eq. (3.67)).

In addition our approach offers some benefits. Rewriting (\ref{RR effective action EM general dipole}) with the notation ${\bf D} \equiv {\bf Q}$ we have  \be
\hS_{EM} = \frac{2}{3}\int\!\!dt \, \hat{\bf D} \cdot \d_t^{3}{\bf D}
\label{hSD}
\ee
This expression applies not only to single charge, but also to a system of charges if only we set \be
{\bf D} \left[ {\bf x}_a \right] := \sum_a q_a\, {\bf x}_a \ee
where the sum is over all the particles in the system. To get the radiation reaction force on any specific charge in the system we need only vary the single object $\hS$  \be
F_{RR,a}^i = \frac{\delta \hS}{\delta {\bf x}_a^i}
\ee

Moreover, the form  (\ref{hSD}) and the Feynman diagram (\ref{RR effective action EM general dipole}) reveal that the dipole appears in the self-force twice: once in an obvious way as ${\bf D}$, the source of the radiation and reaction fields, and a second, less obvious, time as $\hat{\bf D}$ which is the ``target'' coupling through which the reaction field acts back on the charges.  In this sense the source and the target are seen to be connected.

Going beyond the non-relativistic limit, an expression for all the relativistic corrections was given in eq. (3.65) of \cite{BirnholtzHadarKol2013a}.
When expanded, the first relativistic correction includes the electric quadrupole term, the magnetic dipole term, and relativistic corrections to the electric dipole.
These were shown to confirm the expansion of Dirac's formula to next-to-leading order (eq. (3.68) there).

\section*{Conclusions}
\label{section:Conclusion}

Inspired by the (quantum) Closed Time Path formalism, we have shown how dissipative systems can be treated with an action principle in classical contexts.
The explicit algorithm to find the generalized action incorporates field doubling, zone separation and  spherical waves.
They were demonstrated by deriving the generalized (dissipative) action for a classical oscillator attached to a rope (\ref{RR effective action})
and for the Abraham--Lorentz--Dirac EM self-force (\ref{RR effective action EM}).
While these two problems may seem remote from each other, and involve different dimensionality, fields, and sources, the treatment of their radiation follows a very similar path.

\subsection*{Acknowledgments}

We thank B. Kosyakov for encouragement and many helpful comments and B. Remez for commenting on a draft.

This research was supported by the Israel Science Foundation grant no. 812/11 and it is part of the Einstein Research Project "Gravitation and High Energy Physics", which is funded by the Einstein Foundation Berlin.
OB was partly supported by an ERC Advanced Grant to T. Piran.

\appendix
\section{Gauge-invariant spherical waves for electromagnetism}
\label{sec:gauge}

In this appendix we provide details on spherical electromagnetic waves and the definition of $\fieldA, \fieldD$ which are used in the main text and represent the field and source, respectively, of the electric dipole sector.
While it is rather technical in nature, its main purpose is simple: reducing the 3+1\,d EM action (\ref{MaxwellEMAction}) - the action of a 4-vector field in 4d - to the action of a single field in 1+1\,d, in complete analogy with the mass--rope system.
The single field will be labeled $\fieldA(r,t)$, and it will be coupled to the source $\fieldD(r,t)$.
The reduction is performed by first singling out the radial coordinate $r$ from the spherical coordinates and then using the tool of gauge invariant spherical fields, mentioned in Sec. \ref{section:Intro}.
The technicalities themselves are best returned to after a full reading.

To utilize spherical symmetry, we define the basis of symmetric trace free (STF) multipoles $X^{L_\ell}=x^{< k_1}x^{k_1}\cdots x^{k_\ell >}=\( x^{k_1}x^{k_1}\cdots x^{k_\ell}\)^{STF}$\footnote
{ 	
For completeness we mention that there are also divergence-less vector multipoles
$x^L_\Omega\!=\!(\vec{r}\! \, \times\!\! \, \vec\nabla x^L)_\Omega$,
but they do not contribute at leading order at $v/c$ ; see \cite{BirnholtzHadarKol2013a}.
},	
which satisfy
\be
\Delta_\Om x^{L} = -\frac{\ell(\ell+1)}{r^2}\, x^{L} ~~ , ~~
\int x_{L_\ell}x^{L'_{\ell '}}\,d\Om = \frac{4\pi r^{2\ell}}{(2\ell+1)!!} \,\delta_{\ell \ell'} \,\delta_{L_\ell}^{L'_{\ell'}} ~ .
\label{spherical multipoles normalizations}
\ee
By also Fourier-transforming over time, we express the fields and sources as
\bea
A_{t/r}=\int\frac{d \w}{2\pi}\sum_L A_{t/r}^{L \, \w} \, x_L e^{-i \w t} ~&,&~~~
A_{\Omega}=\int\frac{d \w}{2\pi}\sum_L \left( A_{S}^{L \, \w}\, \d_{\Omega} \, x_L + A_{V}^{L \, \w} \,x^L_{\Omega} \right)  e^{-i \w t} \, , \nonumber \\
J^{t/r}=\int\frac{d \w}{2\pi}\sum_{L} J^{t/r}_{L \, \w} \, x^L e^{-i \w t} ~&,&~~~
J^{\Omega}=\int\frac{d \w}{2\pi}\sum_L \left( J^{S}_{L \, \w} \,\d^{\Omega} x^L + J^{V}_{L \, \w} \,x_L^{\Omega} \right)  e^{-i \w t} \, ,
\label{decomposition of EM field and sources}
\eea
where the four general functions comprising the field $A^\mu (t,{\bf x})$ have been replaced by $A^t_{L \, \w}(r),A^r_{L \, \w}(r),A^S_{L \, \w}(r),A^V_{L \, \w}(r)$, which are each 1-dimensional.
The leading order expression in $v/c \ll 1$ (\ref{EOM for charge ALD}), i.e. the Abraham--Lorentz force, requires only the dipole ($\ell=1$) electric components of the field and source\footnote
{ 	
Higher multipole orders (of both the scalar (electric) and vector (magnetic) fields) correspond to relativistic corrections of higher order in $v/c$, and must be included to reproduce the full ALD force \cite{BirnholtzHadarKol2013a}.
},	
i.e. only the 1-dimensional vector fields\footnote
{ 	
We henceforth drop the index $\w$, it will be implied. Note ${\bf A}_{-\w} \!=\! {\bf A}^{*}_{\w}, {\bf J}_{-\w} \!=\! {\bf J}^{*}_{\w}$ since $A_{\mu},J^{\mu}$ are real.
} 
${\bf A}_{t}(r),{\bf A}_{r}(r),{\bf A}_{S}(r)$ and source vectors ${\bf J}^t(r),{\bf J}^r(r),{\bf J}^S(r)$, where
\bea
{\bf J}^t(r)&=&
\frac{3!!}{4\pi r^2}\!\int\!\! \rho_\w({\bf x})\, {\bf x}\, d\Omega=
\frac{3}{4\pi r^2}\!\int\!\!\!\!\int\!\!{dt}\,e^{i \w t} \rho({\bf x},t)\, {\bf x} \,d\Omega ~ ,	\nonumber\\
{\bf J}^r(r) &=&
\frac{3}{4\pi r^2}\!\int\!\! \( \vec{J}_w(\vec{r})\cdot \hat{r}\) {\bf x}\, d\Omega=
\frac{3}{4\pi r^2}\!\int\!\!\!\!\int\!\!{dt}\,e^{i\w t} \( \vec{J}(\vec{r},t)\cdot \hat{r}\) {\bf x} \,d\Omega ~ ,
\label{EM inverse sources}
\eea
and ${\bf J}^S(r)$ is replaced using the equation of current conservation
\be
0=D_\mu {\bf J}^\mu = i \w {\bf J}^t + (\d_r+\frac{3}{r}){\bf J}^r- 2{\bf J}^S.
\label{current conservation}
\ee
Substituting (\ref{decomposition of EM field and sources}) into Maxwell's action (\ref{MaxwellEMAction}) and using (\ref{spherical multipoles normalizations}) we obtain (at leading order)
\bea
S&=&\frac{1}{6}\!\int\!\!\frac{d \w}{2\pi}\!\int\!\!dr\, r^{4}
	\left\{
		\left| i \w {\bf A}_r - \frac{1}{r}(r {\bf A}_t)' \right|^2
		 +\frac{2}{r^2}  \left| i \w {\bf A}_S - {\bf A}_t \right|^2
	\right.
\nonumber \\
	&&~~~~~~~~~~~~~~~~~~
\left.
	-\frac{2}{r^2} \left| \frac{1}{r}({\bf A}_S)' - {\bf A}_r \right|^2
	\!\!\!-\! 4\pi \! \left[ {\bf A}_r\!\cdot\!{\bf J}^{r *} + {\bf A}_t\!\cdot\!{\bf J}^{t *}
						+ 2 {\bf A}_S \!\cdot\! {\bf J}^{S *} + c.c. \right] \right\}
\!,~~~~~~~
\label{EM action spherical}
\eea
where $':=\frac{d}{dr}$.
We notice that ${\bf A}_r$ is an auxiliary field, which means its derivative $A'_r$ does not appear in (\ref{EM action spherical}).
Hence, its EOM is algebraic and is readily solved,
\be
{\bf A}^r = -\frac{1}{\w^2 - \frac{2}{r^2}} \left[\frac{i \w}{r}(r {\bf A}^{t})' + \frac{2}{r^3}(r {\bf A}^{S})' - 4\pi {\bf J}^r \right] ~.
\label{Ar}
\ee
Substituting this algebraic solution for the field back into the action and using (\ref{current conservation}) for the sources, we see that the 1-dimensional action depends in fact on a single gauge invariant 1-dimensional field $\gaugeA$ with a single source $\gaugeD$:
\be
S = \frac{1}{6}\int\frac{d \w}{2\pi} \int\!\! dr\, r^4
	\left[
			\frac{1}{1-\Lambda} \left| \frac{1}{r} (r \gaugeA)' \right|^2
			+\frac{c_s}{r^2} \left| \gaugeA \right|^2
			+4\pi (\gaugeA \!\cdot\! \gaugeD^* +c.c.) \right],~~~
\label{SS}
\ee
where
\be
\gaugeA :={\bf A}_t - i\w {\bf A}_S ~~~~,~~~~
\gaugeD := - {\bf J}^t+\frac{i}{\w r^3} \left( r^3 \frac{\Lambda}{\Lambda-1} {\bf J}^r \right)' \label{SSASSrho} ~ ,
\ee
and $\Lambda:=\frac{\w^2 r^2}{2}$.
For the single charge $q$ with trajectory ${\bf x}_p (t)$, the second summand on the right hand side of $\gaugeD$ (\ref{SSASSrho}) is sub-leading $\(\sim (\w r)^2\)$ and can be dropped, leaving
\be
\gaugeD=-{\bf J}^t=-\frac{3q}{4\pi r^2}\!\! \int\!\!d\Om\, {\bf x}\, \delta({\bf x} - {\bf x}_p) \, \, .
\label{EM source scalar}
\ee
In order to bring the action (\ref{SS}) closer to canonical form we will define a new field $\fieldA$, which is essentially the momentum conjugate to $(r\gaugeA)$, via the transformation
\be
~~\left(\frac{r^2 \fieldA}{3}\right):=\frac{\d L}{\d(r\gaugeA^*)'} \, = \, \frac{r^2 (r\gaugeA)'}{3 (1-\Lambda)}~~~~;~~~~
\left(\frac{r^2 \fieldA}{3}\right)':=\frac{\d L}{\d(\gaugeA^*)} \, = \, \frac{2}{3}(r\gaugeA) \, + \, \frac{4\pi r^3\gaugeD}{3} ~ ,~~
\label{PI S2}
\ee
and by differentiating once more w.r.t $r$ and combining these two equations, we find
\bea
0&=&\frac{r^4}{6}\left(\w^2+\d_r^2+\frac{4}{r}\d_r \right) \fieldA-\fieldD ~ ,
\label{EOM PhiS}
\eea
where
\be
\fieldD=\frac{2\pi r^{2}(r^3 \gaugeD)' }{3}
=-\frac{q r}{2}\!\!\int\!\!d\Omega\, {\bf x} \left[r^2 \delta({\bf x} - {\bf x}_p) \right]' ~ ,
\label{EM source scalar Phi}
\ee
as the new field's equation and source term.

Altogether in this sector the 1-dimensional action is given by
\be
S=\int\frac{d \w}{2\pi} \int\!\! dr\,
	\left[
		\frac{r^4}{6} \fieldA^{\!*} \!\cdot\! \left(\w^2+\d_r^2+\frac{4}{r}\d_r \right) \fieldA
		 -\( \fieldA^{\!*} \!\cdot\! \fieldD + \fieldA \!\cdot\! \fieldD^* \)
	\right] ~ .~~~
\label{EM Action leading2}
\ee
This 1+1\,d action is (to leading order) the EM analog of the mass--rope action, and serves as the starting point for field doubling and solution ((\ref{EM Action leading}) and on).

\section{Bessel functions}
\label{sec:Bessel}

Equation (\ref{Modified Bessel equation}) is a special case of
\be
\[ \del_x^2 + \frac{2\alpha+1}{x}\del_x + 1\] \tilde{b}_\alpha(x) =0 ~,
\label{Modified Bessel equation2}
\ee
for order $\alpha=3/2$.
As described in \cite{BirnholtzHadar2013b} (see eq. B.6), it matches the case of dipole ($\ell=1$) waves in $d=4$ spacetime, as $\alpha=\ell+\frac{d-3}{2}$.
The solutions are given by the origin-normalized Bessel functions
\be
\tilde{b}_\alpha := \Gm(\alpha+1) \, 2^{\alpha} \, \frac{B_\alpha(x)}{x^\alpha} ~ ,
\ee
where $B \equiv \{J,Y,H^\pm\}$ includes Bessel's functions of the first and second kind $J,Y$ and Hankel's functions, $H^\pm=J \pm i\, Y$.
With this definition, $\tilde{j}_\alpha$ behaves smoothly in the vicinity of the origin $x=0$,
\be
\tilde{j}_\alpha (x) = \sum_{p=0}^\infty \frac{(-)^p \, (2\alpha)!!}{(2p)!! (2p+2\alpha)!!} x^{2p} = 1 + \co\(x^2\)~,
\label{Bessel J series2}
\ee
and the Hankel functions' asymptotic form for $x\to\infty$ is given by
\be
\tilde{h}^\pm_\alpha(x) \sim (\mp i)^{\alpha+1/2} \frac{2^{\alpha+1/2} \Gm(\alpha+1)}{\sqrt{\pi}} \frac{e^{\pm i x}}{x^{\alpha+1/2}} ~ .
\label{Bessel H asymptotic2}
\ee
For further details see Appendix A.2 of \cite{BirnholtzHadarKol2013a} or Appendix B.2 of \cite{BirnholtzHadar2013b}. Note the conventions for the subscripts of $\tilde{j}$ are slightly different; we adopt those of \cite{BirnholtzHadar2013b}.


\end{document}